# RR Lyrae Variables in NGC 2808

ANDREA KUNDER[1], PETER B. STETSON[2], MÁRCIO CATELAN[3], PÍA AMIGO[3], ROBERTO DE PROPRIS[1]

(1) NOAO-Cerro Tololo Inter-American Observatory, La Serena, Chile
(2) Dominion Astrophysical Observatory, Herzberg Institute of Astrophysics, National Research Council, Victoria BC, Canada
(3) Pontificia Universidad Católica de Chile, Departamento de Astronomía Astrofísica, Santiago, Chile

## Abstract

NGC 2808 is a unique globular cluster with not only a bimodal-horizontal branch (HB) but also with gaps on the blue horizontal branch. Adequate interpretation of the nature of the detected peculiarities in bimodal and gap clusters is of paramount importance for understanding the nature of the second parameter phenomenon and scenarios for the formation of the Galaxy. Although RR Lyrae variables are HB stars that can provide powerful constraints to models on the origin of bimodal HBs, unfortunately, until recently, only one RR Lyrae variable was known in this cluster. Here we present the first calibrated time-series CCD photometry for newly discovered fundamental mode RR Lyrae variables in NGC 2808, with observations over a range of twenty years. Investigations of RR Lyrae variable stars in this peculiar, bimodal-horizontal branch globular cluster are carried out to account for its formation, and the effects of helium enrichment and differential reddening. The Oosterhoff classification of NGC 2808, which has recently been associated with a previously unknown dwarf galaxy in Canis Major, is also discussed.

## 1. Introduction

NGC 2808 has been dubbed *anomalous* for at least thirty years, when a bimodal distribution of horizontal branch (HB) stars was observed. Today, the details surrounding its HB is still widely debated. Here we begin an observational analysis on one of its HB components, the RR Lyrae instability strip, which to date has not been sufficiently observed.

The HB morphology can differ substantially in the globular clusters (GCs) in our Galaxy (see, for example, Piotto et al. 2002). Some Milky Way globular clusters are known to contain only red HB stars cooler than the RR-Lyrae gap. Other clusters host blue HB stars, even bluer than the canonical end of the HB at ∼35000 K (e.g. Moehler et al. 2004). Then there are clusters with both red and blue HB stars, and hence have HBs that span a wide color range (e.g. M 15, Piotto et al. 2002; Momany et al. 2002; NGC 2808, Sosin et al. 1997). It is well known that the metallicity of a GC influences the HB substantially, but this "first parameter" cannot account for the complex observational picture. An additional "second parameter" is needed in order to interpret the HBs in GCs (Sandage & Wildey 1967; van den Bergh 1967).

Age has been proposed as a second-parameter to explain these different HB morphologies (Dotter et al. 2010, and references therein), but age is unlikely to be the sole explanation (e.g., Gratton et al. 2010). Mass loss (Catelan 2000), primordial abundance variations (D'Antona et al. 2002), stellar rotation (Norris 1983), deep mixing on the red giant branch (RGB; Sweigart 1997), core density (Buonanno et al. 1997), and planetary systems (Soker & Harpaz 2000), among others, are additional parameters which may also play a role in HB morphology. Given all these factors, GCs presenting the second parameter syndrome internally – that is, those with bimodal HBs – are especially valuable to understand the phenomenon (e.g., Rood et al. 1993; Catelan et al. 1998a,b; Corwin et al. 2004).

NGC 2808 is a GC with three main sequence (MS) branches (Piotto et al. 2007), and these branches are likely associated with the complexities of the cluster's HB. It has been put forth by D'Antona et al. (2005) that the three MS populations have three different helium abundances, which could also explain the HB observations. According to their models, the RR Lyrae objects could be the bluest extension of the red clump population, which has a normal $Y$ (D'Antona et al. 2005). They could also be the reddest part of the blue clump, which contains stars slightly helium enhanced ($0.26 < Y < 0.29$). It is the extended blue tail that would consist of the most helium enhanced stars. Other models predict that RR Lyrae stars would be helium enhanced ($Y \sim 0.33$, Lee et al. 2005, see their Fig. 3; $Y \sim 0.3$, Dalessandro et al. 2010, see their Fig. 7). Hence, when comparing theory to observation, it is important to know if the RR Lyrae show signs of enhanced helium.

RR Lyrae were not thought to be a major constituent of NGC 2808 until Corwin et al. (2004) discovered ∼15 RR Lyrae variables close to its crowded core. However, due to extreme crowding, they were unable to place the light curves on a photometrically calibrated scale. But different models for the origin of bimodal HBs require knowledge on the brightness of these pulsating variables. Further, amplitudes of the RR Lyrae variables are indicative of their $T_{eff}$, which in turn can be compared to their period to understand potential "period shifts". The period shift is usually made between the RR Lyrae stars in different globular clusters (Sandage 1981; Sandage et al. 1981; Carney, Storm & Jones 1992; Sandage 1993), and can be used as an indicator for [Fe/H], luminosity, or Oosterhoff-type (e.g., Cacciari et al. 2005). In particular, if

He enhancement is present, large period shifts are clearly expected (see, e.g., Sweigart et al. 1987)

Comparing the period-temperature relations of RR Lyrae variables in bimodal-HB GCs with those in GCs that do not show the second-parameter effect can provide important insight into their HB bimodality. As Corwin et al. (2004) point out, both mass loss variations among RGB stars, and internal age variations within the cluster affects only the mass with which a star will arrive on the zero-age HB (ZAHB). This would cause an RR Lyrae variable to be shifted horizontally in the period-amplitude (PA) plane. In contrast, vertical shifts in the PA plane are indicative of higher luminosities. Therefore, comparing the period shifts with respect to the RR Lyrae in "normal" GCs should be indicative of second parameters that do directly affect HB luminosity (Catelan 1998b; Corwin et al. 2004)

Lastly, the RR0 Lyrae variables position on the PA plane can be used to investigate suggestions that NGC 2808 is part of the Canis Major dwarf galaxy (Crane et al. 2003; Forbes, Strader & Brode 2004). RR Lyrae variables in neighboring dwarf spheroidal galaxies have different PA relations and different Oosterhoff types than the RR Lyrae in the MW (see Catelan 2009 for a review). Hence, the Oosterhoff type of the RR Lyrae in NGC 2808, given by its position in the PA plane, can give insight to if this GC belongs to the MW or if it had its origins in a dwarf spheroidal galaxy.

## 2. Observations and Data Reduction

The CCD imaging observations used here include all the observations of NGC 2808 obtained by Corwin et al. (2004) as well all publicly available NGC 2808 data that we were able to locate; these images are now contained within a private archive maintained by author PBS. Further observations were carried out specifically for this project in Jan 2010 and Jan 2011 using the Y4KCam on the SMARTS 1m telescope. The Jan 2011 has not yet been incorporated into our analysis, so our data thus span a range from 1983 to 2010. There are ∼600 individual CCD images from ∼25 observing runs. These data have been placed by one of the authors (PBS) on the photometric system of Landolt (1992; see Stetson 2000, 2005) The average number of independent observations for a star was 150 in $B$, 200 in $V$, and 50 in $I$. However, due to the severe crowding, only about half of these observations have photometric errors that make them usable ($\sigma_V < 0.1$ mag). Here we concentrate on the $V$ and $I$ observations only, and only the fundamental mode pulsators. Inclusion of the $B$ observations as well as the RR1 variables will be reported on in the future.

## 3. Analysis

All of the Corwin et al. (2004) variables were recovered, except for V21, V22 and V27. From the HST data of NGC 2808, V27 looks like either a tight equal-light binary or a galaxy. There is no star at the position of V21, and the stars closest do not seem to be varying, or at least haven't been measured with sufficient reliability. There is no star at the position of V22, but there does appear to be a star 1.8 arcsec away that may be varying. However, this star has a magnitude that is too bright to be an HB star. V12 is recovered using the following coordinates: RA=09:11:56.15, DEC=-64:50:09.6 (J2000.0), which is 3.3 arcsec away from the published position. We do not have reliable photometry for V19, because from a stacked HST image, V19 is blended with a fainter star and is 0.85 arcsec away from a much brighter (V=14.3) star.

As the majority of the obtained observations are in the $V$-band, the $V$ light curves are used to derive periods. Both an RR Lyrae template fitting program (Layden et al. 1999; Layden & Sarajedini 2000) and the phase dispersion minimization technique (Stellingwerf 1978) are used for period determinations. The best periods were then used to phase the $I$-band light curve to ensure the period phased both light curves equally well. This worked well for most stars, except for the stars that exhibited the largest light curve scatter (e.g., V25). The periods found are within the errors to those published by Corwin et al. (2004). The light curves for the fundamental mode pulsators are provided in Figure 1, and preliminary periods and magnitudes are given in Table 1.

All of the RR0 variables show evidence of light curve modulation, with the exception of V6. Although the Corwin et al. (2004) observations have a rather short time baseline (2 nights in December 2002, and two nights in February 2003), this effect is seen in many of their light curves as well. The light curve modulation is likely due to the Blazhko effect, which was discovered for the first time in 1907 (Blazhko 1907) and appears as a cyclic modulation of shape and amplitude of the light curve. Despite a century of study, the phenomenon is still not understood.

The 5 RR0 Lyrae stars exhibiting the Blazhko behavior are V14, V15, V16, V18, and V25; hence 83% of the NGC 2808 population exhibits the Blazhko behavior. This is a larger percentage than seen in $\omega$ Cen (20% found by Jurcsik et al. 2001), a larger percentage than seen in the field RR0 stars (47% found by Jurcsik et al. 2009), and is likely one of the largest percentages for MW GCs. It has been shown that the Blazhko effect favors short periods (P $\sim$ 0.5 d, Jurcsik et al. 2011; see also Preston 1964). The fact that all the NGC 2808 RR0 Lyrae variables are OoI-type (which have shorter periods on average than OoII-type RR Lyrae variables) is consistent with their result.

Our Blazhko ratio is perhaps similar to that of M107, a GC with a similar metallicity to NGC 2808 and also, like NGC 2808, a blue and red HB (Piotto et al. 2002). Clement & Shelton (1999) find 6 out of their 9 RR0 Lyrae variables have peculiar light curves (67%), which could be attributed to the Blazhko effect, and all of these 9 RR0 Lyrae variables are of OoI-type. If M107 and NGC 2808 had a similar formation mechanism, this would seem to suggest that the Blazhko phenomenon is influenced by evolution of an RR Lyrae variable and how the star was formed. This has also been but forth by Jurksic et al. (2011) from a detailed study of the $(B-V)$ colors of Blazhko RR Lyrae variables.

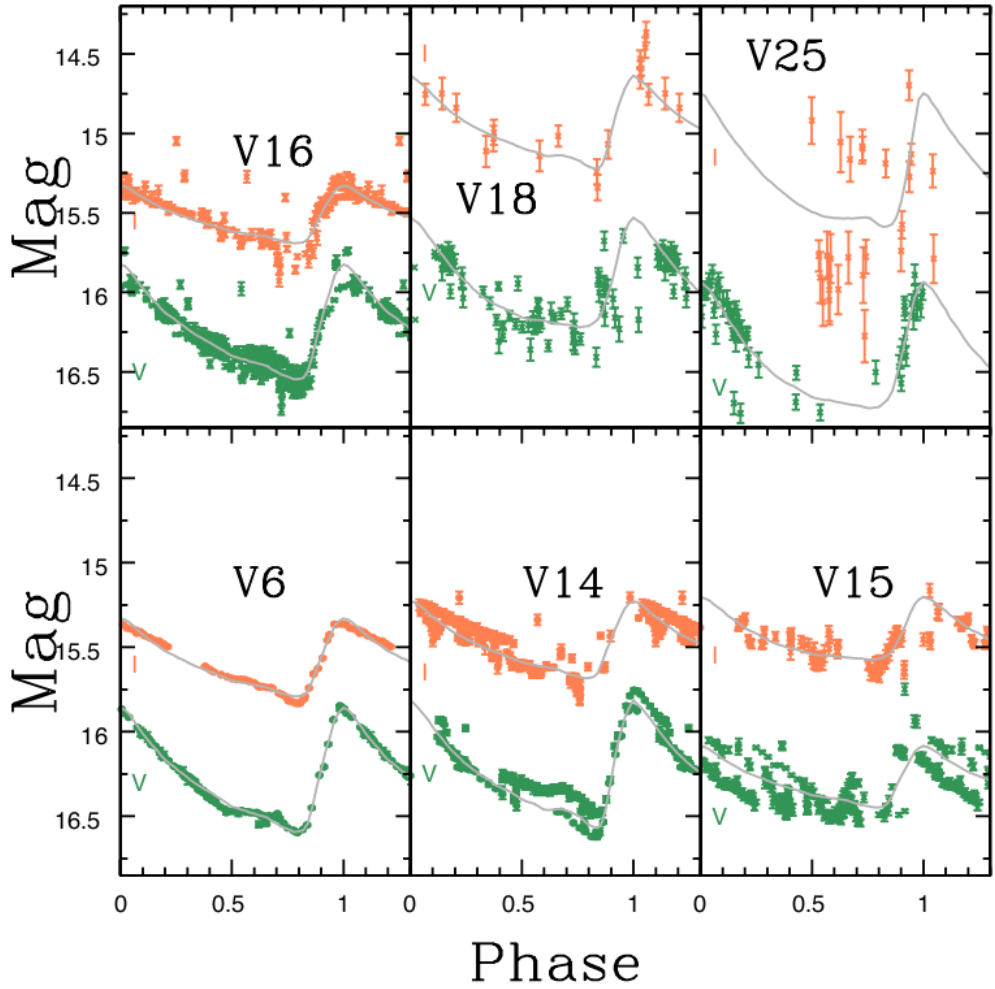

**Figure 1.**— Light curves for the NGC 2808 RR0 Lyrae variables. Lines shown are template fits to the light curves.

There are also some suggestions that the percentage of stars exhibiting the Blazhko effect is correlated with [Fe/H] (Moskalik & Poretti 2003), in a sense that the number fraction increases with decreasing metallicity. However, this GC is more metal-rich than many GCs in the Milky Way. It will be interesting to see how many of the RR1 stars also show evidence of the Blazhko effect, as the RR1 stars have a lower incidence rate of exhibiting the Blazhko effect as compared to RR0 Lyrae stars.

### 3.1. Period-Amplitude Plane

It has often been assumed that the period-amplitude (PA) relation for RR0 variables is a function of metal abundance. Preston (1959) plotted ∼50 field RR0 variables in the period-amplitude plane and demonstrated that the more metal-poor and more metal-rich stars appeared to define two sequences well separated in amplitude. Since then, Clement & Shelton (1999) have found that $V$-amplitude for a given period is not a function of metal abundance, but rather, of the Oosterhoff type.

More recently, Cacciari et al. (2005) compared the PA diagrams for OoI and OoII clusters spanning a wide range of metallicities, and their results show that OoII clusters in particular do not follow a single line in this plane, with the position of the "classic," metal-poor OoII clusters such as M15 and M68 being shifted to shorter periods compared to more metal-rich OoII clusters (see also Corwin et al. 2008).

Although it is unclear what governs the Oosterhoff type of an RR Lyrae star, theoretically, OoII-type RR0 variables have higher luminosities, which can be caused by a number of factors such as enhanced helium abundance (Bono et al. 1997; Busso et al. 2007; Caloi & D'Antona 2007), and also by evolution off the ZAHB (Clement & Shelton 1999). Observationally, OoII-type stars seem to have a period shift of about $-0.06$ with respect to OoI-type stars, (Cacciari et al. 2005, see also Figure 2), although there are some "classically" OoII-type GCs with RR Lyrae variables that do not follow the OoII line well (such as M15, see Corwin et al. 2008).

Figure 2 shows the PA diagram of the RR0 Lyrae stars in NGC 2808. Also plotted are the NGC 1851 RR Lyrae variables from Walker (1998). It is striking that all of the NGC 2808 stars fall along the OoI line. NGC 1851 is a GC similar to NGC 2808 in that they both have an [Fe/H] $\sim -1.2$ and both host multiple populations. The PA relationship of the RR Lyrae in this cluster is strikingly different, giving credence to suggestions that an RR Lyrae variables position in the PA diagram is a function of Oosterhoff type and not of [Fe/H] (Clement & Shelton 1999, Kunder & Chaboyer 2009).

The two GCs NGC 6388 and NGC 6441 have extended blue horizontal branches with red components, similar to the HB of NGC 2808. However, the RR0 variables in these two clusters have abnormally long periods, even longer than typical OoII-type variables. It has been shown that different initial He contents could account for the morphology of the HB in both of these clusters (Busso et al. 2007). As the NGC 2808 RR Lyrae variables have periods and amplitudes that are not consistent with NGC 6388

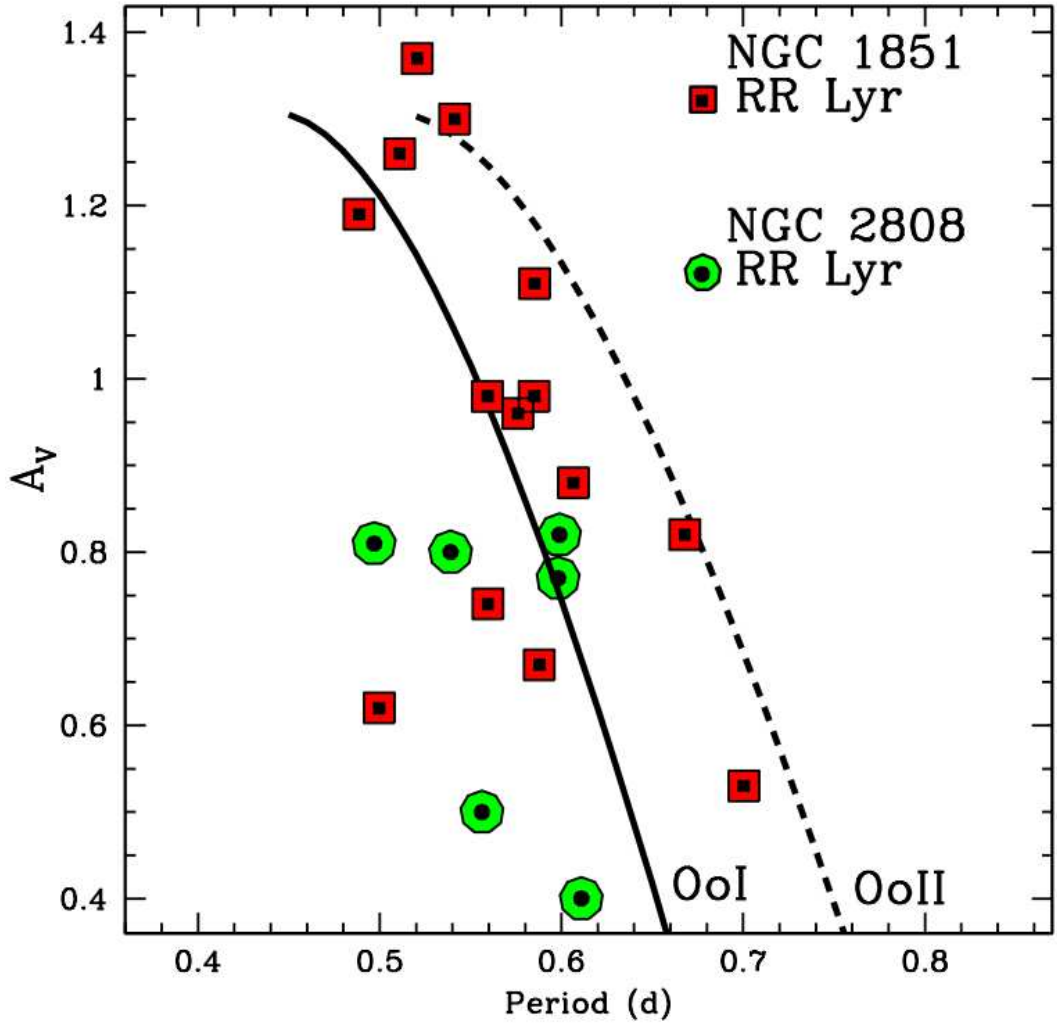

**Figure 2.**— The period-amplitude (PA) diagram for the RR0 Lyrae in NGC 1851 and NGC 2808. The OoI line was derived by Cacciari et al. (2005) from a least-squares fit to the principal sequence of regular RR0 Lyrae stars in the OoI cluster M3. The OoII line is shifted by $\Delta\log P = -0.06$, and was shown to correspond approximately to the mean location traditionally assigned to OoII variables.

or NGC 6441, it is unlikely that these stars are He enriched or share a similar formation history.

Figure 2 shows that the Oo-type of NGC 2808 is OoI, and there are no suggestions that it is an Oo-intermdiate GC. Hence, the RR Lyrae variables in NGC 2808 have properties that are not consistent with the RR Lyrae properties in GCs of neighboring dwarf spheroidal galaxies.

Blazhko variables may introduce scatter into the period-amplitude diagram, because the PA relations for OoI and OoII variables apply when they are at maximum ampli-tude. As shown above, almost all the RR0 Lyrae variables in NGC 2808 show signs of Blazhko behavior. Especially for V18 and V25, where the light curve is not well sampled, we can not rule out that the amplitude measured is at maximum. However, the NGC 2808 RR Lyrae variables would require a ∼0.4 mag offset in their $V$-amplitudes to shift them from being on the OoI-locus to OoII-locus. Our data span a long time baseline (∼20 years), and from visual inspection of the light curves (see Figure 1), it seems unlikely that the Blazhko amplitudes are changing by that much.

### 3.2. Reddening

The interstellar extinction toward NGC 2808 is quite large for a halo GC (E(B−V) = 0.19 mag; Bedin et. al 2000) and it is thought that a small amount of differential reddening exists ($\Delta E(B-V) = 0.02$, Piotto et al. 2007). Here minimum light color of RR0 Lyrae variables are used to investigate the reddening of NGC 2808.

Mateo et al. (1995) used an 11 star sample and Guldenschuh et al. (2005) used 16 RR0 Lyrae variables to find that $E(V-I)$ depends upon only minimum $(V-I)$ color. Guldenschuh et al. (2005) concluded that the intrinsic minimum light color of RR0 Lyrae variables is $(VI)_{min,0} = 0.58 \pm 0.02$, with very little dependence on period or metallicity for periods between 0.39 and 0.7 days and metallicities in the range $-3 \leq [\text{Fe/H}] \leq 0$. Similar results have been confirmed by Kunder, Chaboyer & Layden 2010 in the $V$ and $R$ passbands.

The $(V-I)$ color at minimum light is found by averaging the color over the phase range 0.5-0.8, and this value is listed in the last column of Table 1. The error in the mean $(V-I)_{min}$ is 0.02 mag. We do not have sufficient $I$-band photometry for the determination of $(V-I)_{min}$ for V18 or V25. The average E(V−I) is 0.25 ± 0.02, or assuming a standard reddening law, E(B−V) is 0.19 ± 0.02. This is consistent with both the reddening found by Bedin et al. (2000) and the reddening value adopted in the Harris (1996) GC catalog. It is slightly smaller than the Schlegel et al. (1998) estimate of $E(B-V) = 0.23$.

## 4. Conclusions

The first calibrated $V$-band light curves of the RR0 Lyrae variables in NGC 2808 are presented. A substantial percentage of them, 83%, exhibit the Blazhko effect, and all of the RR0 Lyrae variables appear to be OoI-type. The RR0 Lyrae variables position on the period-amplitude diagram reveals that these stars are not helium enhanced, and do not have abnormally large luminosities. Hence, recent models to explain the HB of NGC 2808 seem to overestimate the level of He enhancement for the RR Lyrae variables in this cluster. The RR0 Lyrae variables have periods and amplitudes that are inconsistent with NGC 2808 being part of the Canis Major dwarf galaxy. From the RR0 Lyrae minimum light colors, the reddening toward NGC 2808 is found to be $E(V - I) = 0.25 \pm 0.02$, in agreement with other reddening estimates of this cluster.

TABLE 1
Galactic Bulge Star Atmosphere Parameters

| Star | Period(d) | $<V>$ | $<I>$ | $A_V$ | $(V-I)_{min}$ |
|------|-----------|-------|-------|-------|---------------|
| V6   | 0.5389687 | 16.27 | 15.59 | 0.80  | 0.79          |
| V14  | 0.5989    | 16.25 | 15.51 | 0.78  | 0.82[b]       |
| V15  | 0.6109    | 16.30 | 15.45 | 0.40  | 0.86[b]       |
| V16  | 0.6052    | 16.23 | 15.53 | 0.77  | 0.80[b]       |
| V18  | 0.5562    | 16.00 | 14.97 | 0.50  | –[b]          |
| V25  | 0.5156    | 16.43 | –     | 0.81  | –[b]          |

[b] Blazhko

**Acknowledgments.** Support for M.C. and P.A. is provided by the Chilean Ministry for the Economy, Development, and Tourism's Programa Inicativa Científica Milenio through grant P07-021-F, awarded to The Milky Way Millennium Nucleus; the BASAL Center for Astrophysics and Associated Technologies (PFB-06); the FONDAP Center for Astrophysics (15010003); and Proyecto Fondecyt Regular #1110326. P.A. acknowledges additional support from SOCHIAS, PUC-DAA, and MECESUP.

## References

Bedin, L. R., Piotto, G., Zoccali, M., Stetson, P. B., Saviane, I., Cassisi, S., & Bono, G. 2000, A&A, 363, 159
Blazhko, S. 1907, Astr. Nachr., 175, 325
Bono, G., Caputo, V., Castellani, V., & Marconi, M. 1997a, A&AS, 121, 327
Busso, G. et al. 2007, A&A, 474, 105
Buonanno, R., Corsi, C., Bellazzini, M., Ferraro, F. R., & Fusi Pecci, F. 1997, AJ, 113, 706
Cacciari, C., Corwin, T. M., & Carney, B. W. 2005, AJ, 129, 267
Caloi, V. & D'Antona, F. 2007, A&A, 463, 949
Carney, B. W., Storm, J., & Jones, R. V. 1992, ApJ, 386, 663
Catelan, M., Borissova, J., Sweigart, A. V., & Spassova, N. 1998a, ApJ, 494, 265
Catelan, M., Sweigart, A. V., & Borissova, J. 1998b, in A Half Century of Stellar Pulsation Interpretation: A Tribute to Arthur N. Cox, ed. P. A. Bradley, & J. A. Guzik (San Francisco: ASP), ASP Conf. Ser., 135, 41
Catelan, M. 2000, ApJ, 531, 826
Catelan, M., 2009, Ap&SS, 320, 261
Clement, C.M. & Shelton, I. 1999, ApJ, 515, L85
Corwin, M.T. et al. 2004, A&A, 421, 667
Corwin, T. M., Borissova, J., Stetson, P. B., Catelan, M., Smith, H. A., Kurtev, R., & Stephens, A. W. 2008, AJ, 135, 1459
Crane, J.D. et al. 2003, ApJ, 594, L119
Dalessandro, E., Salaris, M., Ferraro, F. R., Cassisi, S., Lanzoni, B., Rood, R. T., Fusi Pecci, F., Sabbi, E. 2011, MNRAS, 410, 694
D'Antona, F., Caloi, V., Montalban, J., Ventura, P., & Gratton, R. 2002, A&A, 395, 69

D'Antona, F., Bellazzini, M., Caloi, V., Pecci, F. Fusi, Galleti, S. & Rood, R. T. 2005, ApJ, 631, 868
Dotter, A. et al. 2010, ApJ, 708, 698
Forbes, D.A., Strader, J. & Broode, J.P. 2004, AJ, 127, 3394
Gratton, R.G., Carretta, E., Bragaglia, A., Lucatello, S., & D'Orazi, V. 2010, arXiv:1004.3862
Guldenschuh, K. A. et al. 2005, PASP, 117, 721
Jurcsik, J., Clement, C., Geyer, G.H., & Domsa, I. 2001, AJ, 121, 951
Jurcsik, J., et al. 2009, MNRAS, 400, 1006
Jurcsik, J., et al. 2011, MNRAS, 411, 1763
Kunder, A.M., & Chaboyer, B. 2009, AJ, 138, 1284
Kunder, A.M., Chaboyer, B. 2009, & Layden, A. 2010 AJ, 139, 415
Landolt, A.U. 1992, AJ, 104, 340
Layden, A.C., Ritter, L.A., Welch, D.L. & Webb, T. 1999, ApJ, 117, 1313
Layden, A.C. & Sarajedini, A. 2000, ApJ, 119, 1760
Mateo, M., Udalski, A., Szymanski, M., Kaluzny, J., Kubiak, M., & Krzeminski, W. 1995, AJ, 109, 588
Moehler, S., Sweigart, A. V., Landsman, W. B., Hammer, N. J., & Dreizler, S. 2004, A&A, 415, 313
Momany, Y., Piotto, G., Recio-Blanco, A., et al. 2002, ApJ, 576, L65
Moskalik, P., & Poretti, E. 2003, Astron. Astrophys., 398, 213.
Norris, J. 1983, ApJ, 272, 245
Piotto, G., King, I. R., Djorgovski, S. G., et al. 2002, A&A, 391, 945
Piotto, G. et al. 2007, ApJ, 661, L53
Preston, G.W. 1959, ApJ, 130, 507
Preston, G.W. 1964, Ann. Rev. A&A, Vol 2., 23
Rood, R. T., Crocker, D. A., Fusi Pecci, F., et al. 1993, in The Globular Cluster-Galaxy Connection, ed. G. H. Smith, & J. P. Brodie (San Francisco: ASP), ASP Conf. Ser., 48, 218
Sandage, A., & Wildey, R. 1967, ApJ, 150, 469
Sandage, A. 1981, ApJ, 248, 161
Sandage, A. 1993, AJ, 106, 687
Sandage, A., Katem, B., & Sandage, M. 1981, ApJS, 46, 41
Schlegel, D. J., Finkbeiner, D. P., & Davis, M. 1998, ApJ, 500, 525
Soker, N., & Harpaz, A. 2000, MNRAS, 317, 861
Sosin, C., Dorman, B., Djorgovski, S. G., et al. 1997, ApJ, 480, L35
Stellingwerf, R.F. 1978, ApJ, 224, 953
Stetson, P.B. 2000, PASP, 112, 925
Stetson, P.B. 2005, PASP, 117, 563
Sweigart, A. V. 1997, ApJ, 474, L23
Sweigart, A. V., Renzini, A., & Tornambè, A. 1987, ApJ, 312, 762
van den Bergh, S. 1967, AJ, 72, 70
Walker, A.R. 1998, AJ, 116, 220